\documentclass[pre,floatfix,twocolumn, 10pt]{revtex4-1} 
\usepackage{graphicx}
\usepackage{amssymb}
\usepackage{natbib}
\usepackage{dcolumn}
\usepackage{bm}
\usepackage{color}
\usepackage[colorlinks=true, linkcolor=blue, citecolor=blue]{hyperref}
\usepackage{subfigure}
\usepackage{graphicx,psfrag,xspace}
\usepackage{color}
\usepackage{amssymb,amsfonts,amsmath}
\usepackage{setspace}

\begin{document}

\author{A. B. Kolton and E. A. Jagla}
\affiliation{Centro At\'omico Bariloche, Instituto Balseiro, 
Comisi\'on Nacional de Energ\'ia At\'omica, CNEA, CONICET, UNCUYO,
Av. E. Bustillo 9500
R8402AGP S. C. de Bariloche, 
R\'io Negro, Argentina}

\title{Thermally rounded depinning of 
an elastic interface on a washboard potential}

\begin{abstract}
The thermal rounding of the depinning transition of an elastic interface sliding on a 
washboard potential is studied
through analytic arguments and very accurate numerical simulations.
We confirm the standard view that well below the depinning threshold
the average velocity
can be calculated considering thermally activated nucleation 
of defects.
However, we find that the straightforward extension of
this analysis
to near or above the depinning threshold does not fully 
describe the physics of the thermally assisted motion.
In particular, we find that exactly at the depinning point
the average velocity does not follow a pure power-law 
of the temperature as naively expected by 
the analogy with standard phase transitions 
but presents subtle logarithmic corrections. 
We explain the physical mechanisms behind these corrections
and argue that they are non-peculiar collective 
effects which may also apply to the 
case of interfaces 
sliding on uncorrelated disordered landscapes. 
\end{abstract}

\maketitle

\section{Introduction}

The depinning transition of elastic interfaces is a paradigmatic example of an out-of-equilibrium critical phenomenon.
Its study is relevant for modeling diverse extended physical systems embedded in a quenched pinning potential. 
Often the pinning landscape acting in the interfaces is disordered such as in driven ferromagnetic~\cite{Ferre2013} and ferroelectric~\cite{Kleemann2007,Paruch2013} domain walls, tension driven cracks~\cite{Bonamy2008,Ponson2009,LePriol2020}, displacement of contact lines of liquid menisci~\cite{Joanny1984,moulinet2004,ledoussal2009} or earthquakes~\cite{jagla2010,jagla2014}. 
In other cases the pinning landscape can be highly correlated or even periodic, 
such as the potential energy of the superconducting phase difference 
in long current driven Josephson junctions~\cite{tinkham2004introduction}, or as in the 
case of field driven domain walls in artificial pinning potentials~\cite{marconi2008,marconi2011,metaxas2013}.

The basic phenomenology of depinning consists of an elastic manifold with an overdamped motion that interacts with a quenched potential energy landscape that tends to trap the interface in configurations in which the potential energy is minimized. When a uniform external driving force is applied, the
interface remains pinned in a local minimum of the tilted energy potential if the amplitude of the driving force is below some well defined threshold, whereas otherwise it sets in a steady-state of motion with a well defined average velocity. 
The threshold value of the applied driving force defines the depinning force of the system $f_c$. 
When the driving force is slightly above the depinning threshold the velocity of the interface is expected to grow as a power law of the excess driving above the threshold value. This is just one indication that the kind of phenomenon occurring near $f_c$ can be characterized as a (non-equilibrium) phase transition with critical properties~\cite{Fisher1998,Kardar1998,brazovskii2004}.

The existence of a sharp depinning transition as a function of the driving force of an elastic interface depends crucially on the fact that thermal effects are negligible. If thermal fluctuations are important, then the depinning transition is smeared out, as for any finite applied force the interface can eventually jump forward via thermally activated events over energy barriers, and hence the average velocity becomes different from zero for any non-zero driving force. It has been proposed that the effect of a small temperature on the depinning transition can be accounted for through an appropriate generalization of the scaling theory used at $T=0$. In this respect, 
it has been suggested, either following a naive analogy with standard equilibrium phase transitions or by phenomenological nucleation theory arguments, that the effect of temperature on the depinning transition can be characterized by the value of a ``thermal rounding'' exponent $\psi$, that describes the average velocity $v$ right at $f_c$ as a function of temperature, namely $v(f_c,T)\sim T^\psi$. 
Regarding the velocity as the order parameter, the force as the control parameter and the temperature as a ``symmetry-breaking field destroying the pinned phase'' \cite{Fisher1985} such scaling proposal for the rounded depinning transition is analogous to the scaling with field $H$ of the equilibrium Ising model magnetization $M$ at the critical temperature $T_c$, $M(T_c) \sim H^{1/\delta}$ with $\delta>0$, to cite the simplest example. The precise determination of the value of $\psi$ has proven to be quite tricky however~\cite{Bustingorry2012,purrello2017}, its universality questioned~\cite{Middleton1992b}, and there is not yet a rigorous proof that the naive thermal rounding scaling theory is 
even consistent, in contrast with the zero temperature dynamics~\cite{ledoussal2002,brazovskii2004,rosso2007,ledoussal2009b} and the subthreshold creep dynamics
~\cite{chauve2000,kolton2006,kolton2009,Ferrero2013}.

In order to advance in the study of the thermal rounding of depinning-like transitions, we concentrate here in a case in which the zero-temperature limit provides an almost trivial result for the flux curve, and where the effect of temperature can be treated in a very accurate if not rigorous manner. 
This is the case of an elastic manifold evolving on a periodic pinning potential, the same 
for all individual sites of the elastic manifold, also known as a ``washboard potential''. 
The model is on the other hand the celebrated
overdamped Sine-Gordon dynamical model which has been used to model many different physical phenomena,
such as the motion of dislocations in the Peierls potential of a crystal~\cite{hull2011introduction} 
overdamped coupled pendula \cite{Buttiker1981}, the equilibrium roughening transition~\cite{chaikin95}, crystal-growth~\cite{hwa1991,barabasi_stanley_1995}, vortex matter in layered superconductors~\cite{blatter1994,nattermann2000,Giamarchi2002,ledoussal2010}, 
forced soliton gases~\cite{Bennett1981} and overdamped long Josephson junctions 
driven by an external current~\cite{tinkham2004introduction}. 
The Sine-Gordon model is closely related to the 
the Frenkel-Kontorova~\cite{braun2013frenkel} 
and Prandtl-Tomlinson models relevant for nanotribology~\cite{popov2012prandtl},
and it may be also used to model
the dynamics of the internal degrees of freedom of an extended magnetic domain wall describing the 
axial rotation of the local magnetization vector, relevant 
for spintronics~\cite{tatara2004,lecomte2009,barnes2012}. 
A proper understanding of the depinning transition of the Sine-Gordon model 
per se is hence also very important, as it is for instance related to the onset of 
rotation of torque driven coupled pendula, the onset of dissipation 
in superconducting systems such as Josephson junctions or vortex systems or to the Walker 
breakdown in magnetic domain wall systems.

In this paper we show, both through analytic arguments and very accurate numerical simulations, that the effect of temperature at the depinning transition in this simple extended model cannot be accounted for by a simple one-parameter scaling, and that it involves the appearance of subtle logarithmic corrections not precluded by any of the standard arguments made so far for the thermal rounding of the depinning transition. We thus expect that this qualitative behavior is not peculiar of the model, but applies for instance to the more standard and complicated case of uncorrelated disorder.

\section{Model}

Consider an elastic interface  (with short range interactions) in $d$ spatial dimensions, characterized by its position $h(r)$. The interface feels the effect of an underlying periodic potential $V(h)$, and an external force $f$. The dynamical equations of the system are \cite{Buttiker1981}

\begin{equation}
\frac{\partial h(r,t)}{\partial t}=-\frac{dV(h)}{dh} +\nabla ^2 h+f +\sqrt T \eta(t,r)
\label{eq0}
\end{equation}
where temperature has been introduced through the use of a standard Langevin formalism, with the white noise $\eta$ characterized by
\begin{eqnarray}
\langle \eta (t,r)\rangle =0,\;\;
\langle \eta (t,r)\eta (t',r')\rangle =2\delta(t-t')\delta^d(r-r')
\end{eqnarray}
At $T=0$ the dynamics of the system greatly simplifies, as the interface becomes flat 
\footnote{Even if the interface is started in an arbitrary $h(r)$ configuration, it gradually flattens, 
and for $t\to\infty$ the interface becomes flat if $T=0$.}, and its global position $h$ follows the one-particle equation 
\begin{equation}
\frac{\partial h}{\partial t}=-\cos(h) +f 
\end{equation}
(from now on we specialize to a potential of the form $V(h)=\sin(h)$).
For $f<f_c=1$ the interface does not move, whereas for $f>f_c$ there is a finite average velocity.
For $f$ slightly above $f_c$, the velocity 
$v \equiv \langle \partial_t h \rangle$ 
scales as
\begin{equation}
v\sim \sqrt{f-f_c}
\end{equation}
which defines the flow exponent $\beta$ (from $v\sim (f-f_c)^\beta$) as $\beta=1/2$.
At finite temperature this sharp continuous transition is smoothed and, 
at variance with the peculiar $T=0$ case, 
the problem becomes a non-trivial collective problem.
In the following sections we discuss the thermally 
activated dynamics below the depinning threshold 
for $|f-f_c|\ll f_c$, and then the subtle 
$f=f_c$ case at finite temperature.

\section{Activated dynamics scaling near the depinning threshold}
\label{sec:activated}
The finite temperature activation dynamics described by Eq.(\ref{eq0}) 
well below $f_c$ was studied in Ref.\cite{Buttiker1981} and also in 
Ref.~[\onlinecite{ettouhami}] using renormalization group methods. 
Here will follow an approach that uses dimensional analysis mainly.
Our aim is to calculate the value of $v$ for finite temperature, and for $f$ very 
close to $f_c=1$.
If $f$ is only slightly below one,  and $T$ is very small ($T\ll (f_c-f)$) the dynamics is dominated by
thermal activation events in which patches of the interface (of linear size $l_0$ to be determined below) advance a definite spatial amount. These patches then grow in size deterministically.

We consider a $d$ dimensional system with periodic boundary conditions at very low temperature, and assume
that  we start with 
a nearly flat 
interface resting in a local minimum of the tilted potential $[\sin(h)-f h]$, 
with $f=f_c-\varepsilon$, and $\varepsilon \ll 1$. 
Then we can approximate the potential by
$\varepsilon h-h^3/6$ near the transition point
\footnote{We are implicitly assuming here that $V(h)-f h$ 
is quadratic around their minima. For a more general 
situation see the Appendix, Sec.\ref{gralnormalform}}. 
The local energy minimum is thus located at 
$h=-\sqrt{2\varepsilon}$, the maximum at $h=+\sqrt{2\varepsilon}$.
Using this expansion in Eq. (\ref{eq0}) the resulting dynamical 
equations that will describe the escape from the energy minimum can be written in a normalized form as
\begin{equation}
\frac{\partial h(r,t)}{\partial t}=\frac{h^2}2 -\varepsilon +\nabla ^2 h+\sqrt T \eta(t,r).
\label{eq1}
\end{equation}
We first calculate the rate of nucleation of defects $R$ (that make the interface advance) per unit of time and unit of volume of the system, in a system with spatial extension $L$. 
$R$ will depend on the two parameters $T$ and $\varepsilon$ present in Eq. (\ref{eq1}) and also on $L$,
i.e., we can write $R(T,\varepsilon,L)$. First we sketch a scaling analysis that allows to reduce the three-parameter dependence of $R$ to a two-parameter dependence.
Suppose we know the value of $R$ for given values of $T$, $\varepsilon$ and $L$.
Then we scale all variables and parameters in Eq. (\ref{eq1}) according to the following table, 
\begin{eqnarray}
\varepsilon &\to& \tilde \varepsilon\equiv k \varepsilon\\
h &\to& \tilde h \equiv k^{1/2}h\\
t &\to& \tilde t \equiv   k^{-1/2} t\\
r &\to& \tilde r \equiv k^{-1/4} r\\
L &\to& \tilde L \equiv k^{-1/4} r\\
T &\to& \tilde T \equiv k^{(6-d)/4}  T
\label{eq:scaling}
\end{eqnarray}
where $k$ is an arbitrary scaling factor. It is readily verified that tilde variables satisfy an equation  formally identical to the original one. 
The above scaling means that the number of activation events in corresponding time and space intervals are equal for the original
and the scaled equation. In concrete, 
\begin{equation}
R(T,\varepsilon,L) [t] [r]^d=R(\tilde T,\tilde \varepsilon,\tilde L) [\tilde t] [\tilde r]^d
\end{equation}
or
\begin{equation}
R(T,\varepsilon,L) =R(k^{(6-d)/4}T,k\varepsilon, k^{-1/4} L) k^{-(d+2)/4}
\end{equation}
Since $k$ is arbitrary we can choose $k\sim 1/\varepsilon$ to obtain
\begin{equation}
R(T,\varepsilon,L) =\varepsilon ^{(d+2)/4} R\left(\frac{T}{\varepsilon^{(6-d)/4}},1,L \varepsilon^{1/4}\right) 
\label{eq:generalactivationrate}
\end{equation}
Alternatively, 
and assuming for simplicity the large system size limit 
($L \varepsilon^{1/4} \gg 1$)
we can accommodate the previous expression as
\begin{eqnarray}
R(T,\varepsilon) =T^{\frac{d+2}{6-d}} \mathcal F\left( \frac{\varepsilon^{(6-d)/4}}{T} \right) 
\label{r}
\end{eqnarray}
where we have dropped the $L$ dependence and defined the unknown function $\mathcal F$ as
\begin{equation}
\mathcal F(x) \equiv x^{\frac{d+2}{6-d}} R(1/x,1,\infty)
\end{equation}
Eq. (\ref{r}) explicitly gives the two parameter form of $R(T,\varepsilon)$ in terms of an  unknown function $\mathcal F$ of a single variable. 
While Eq. (\ref{eq:generalactivationrate}) shows the finite size scaling effects,
the existence of $\mathcal F$ guarantees a well defined thermodynamic limit 
for the nucleation rate, when $L \varepsilon^{1/4} \gg 1$ or $L T^{1/(6-d)} \gg 1$.

The combination $\varepsilon^{(6-d)/4}/T$ in the argument of $\mathcal F$ suggests that $\varepsilon^{(6-d)/4}$ is actually a relevant energy scale of the problem and  
thus we will denote $\alpha=(6-d)/4$ as the ``energy exponent''.
In fact, its physical meaning can be unveiled by a simple variant of the 
``droplet'' argument~\cite{Langer1968,blatter1994}. 
Suppose we want to estimate what is the optimal linear size $l_0$ 
of a patch of the surface to jump the energy barrier implied by 
the force density term $h^2/2-\varepsilon$ in Eq. (\ref{eq1}). 
Assuming simple excitations, solely characterized by its linear size $l_0$ and 
displacement $h$, the additional elastic energy of order
$(h/l_0)^2 l_0^d$ must be added to the potential energy 
$(\varepsilon h - h^3/6)l_0^d$,  
yielding the patch energy near $f_c$,
\begin{equation}
E(h,l_0) \sim (\varepsilon h - h^3/6)l_0^d + (h/l_0)^2 l_0^d/2
\end{equation}
For any $0<d<6$ the excitation energy $E(h,l_0)$ 
has an extremum at
$l_0^* \sim \varepsilon^{-1/4}$ 
and $h^* \approx \sqrt{\varepsilon}$, yielding 
the exact scaling result
\begin{equation}
E^*\equiv E(h^*,l_0^*) \sim \varepsilon^{(6-d)/4}.
\end{equation}
This confirms the physical connection with Eq. (\ref{r}).
For $d<2$ such extremum is a saddle point and $E^*$ is the minimal barrier 
to advance forward.
The optimal 
size $l^*_0 \sim \varepsilon^{-1/4}$ 
is such that the small ($l_0 \ll l^{*}_0$) 
frequently activated patches are futile 
(i.e. they are quickly reversed) while large enough
($l_0 > l^{*}_0$) patches trigger irreversible 
forward jumps of the whole segment.
This physical argument also makes clear that the function $\mathcal F$ in Eq. (\ref{r}) will contain 
a dominant factor $\exp(-C\varepsilon^{(6-d)/4}/T)$ corresponding to an Arrhenius factor 
for the activation of these kind of optimal patches, 
provided $T \ll \varepsilon^{(6-d)/4}$ and under the assumption that the considered segment size 
is larger than $l_0$ \footnote{If the size of the optimal patch becomes of the order of the 
interface size a dimensional crossover is expected towards the zero dimensional or single particle 
result $E_{max} \to \varepsilon^{3/2}$}.

All the previous scaling analysis can be presented also for a 
non-quadratic force minimum, replacing $h^2 \to h^{\gamma}$ 
in the rhs of Eq. (\ref{eq1}), yielding 
(in the limit of large sizes $L T^{\frac{\gamma-1}{2 \gamma -\gamma  d+d+2}}\gg 1$, see Appendix Sec.\ref{gralnormalform}), 
\begin{equation}
R(T,\varepsilon) = T^\sigma 
{\mathcal F}(\varepsilon^\alpha/T)
\label{eq:Fscalinggamma}
\end{equation}
with 
\begin{eqnarray}
\sigma = {\frac{(\gamma -1) (d+2)}{2 (\gamma +1)-(\gamma -1) d}} \\
\label{eq:Fscalinggamma2}
\alpha = {2-\frac{(2+d)(\gamma-1)}{2\gamma}}
\label{eq:Fscalinggamma3}
\end{eqnarray}

Eq.(\ref{eq:Fscalinggamma}) reduces to Eq. (\ref{r}) for $\gamma=2$. 
In particular, Eq.(\ref{eq:Fscalinggamma3}) 
it generalizes the energy exponent $\alpha$.
The above results are valid for estimating the 
thermally activated decay rate of an initially flat 
segment of the interface by the production of a 
single defect. We now analyze in more detail the 
simplest cases, namely the particle and the elastic string, 
keeping the standard $\gamma=2$ for simplicity.

\subsection{Single Particle}
For a single particle ($d=0$) each activation event represents the jump over one barrier, and leads to the advance of the particle by a finite amount $2\pi$. This means that the velocity in the single particle case will follow the scaling:
\begin{equation}
v_{d=0}(\varepsilon>0, T) \sim R_{d=0} \sim T^{1/3}\mathcal F(\varepsilon^{3/2}/T) 
\label{vd0}
\end{equation}

The explicit form of the function $\mathcal F$ in Eq. (\ref{vd0}) and then the form of $v$ is in fact well known in the limit $\varepsilon>0$, $T\ll \varepsilon^{3/2}$, which is the thermally activated regime. 
This corresponds to the Kramers problem of escape over a barrier.
The velocity is simply proportional to the inverse of the escape time of a thermal particle in the potential well 
$\varepsilon h-h^3/6$.
Kramers' formula applied to this case provides 

\begin{equation}
v_{d=0}(T\ll \varepsilon^{3/2})= \sqrt{\varepsilon}\exp \left(- \frac{4\sqrt{2}}{3} \frac{\varepsilon^{3/2}}{T} \right)
\end{equation}
This expression satisfies the scaling expression Eq. (\ref{vd0}).

In the present single particle case the scaling argumentation can be extended to negative (but small) $\varepsilon$, meaning $f$ slightly above the critical value $f_c=1$, since in this case the dynamics is also dominated by the bottlenecks near the points where $v$ is very small. This means that Eq. (\ref{vd0}) can also be used for $\varepsilon<0$. In this case, there is a finite limit for the velocity as $T\to 0$, and for this to be the case $f(-x)\sim (-x)^{1/3}$ for $x\to \infty$, leading to 
\begin{equation}
v_{d=0}(\varepsilon<0, T=0)\sim |\varepsilon|^{1/2}
\end{equation}
which is the expected result. 
Eq. (\ref{vd0}) used at $\varepsilon=0$ also indicates that 
for a single particle $v_{d=0}(\varepsilon=0, T)\sim T^\psi$, with 
a well defined thermal rounding exponent $\psi=1/3$~\cite{ambegaokar1969,bishop1978,purrello2017}.

\subsection{Elastic String}
We now analyze the case $d=1$, corresponding to an elastic string.
Eq. (\ref{r}) becomes in this case
\begin{equation}
R_{d=1}(T,\varepsilon) =T^{3/5} \mathcal F(\varepsilon^{5/4}/T) 
\end{equation}
In $d=1$ the relation between $R$ and the velocity of the interface can be worked out as follows: $R$ represents the rate of creation of kink/anti-kink pairs. Each kink or anti kink moves at a velocity $\pm c$ \footnote{The elastic force at the center of the kink is approximately zero, 
so it is only acted by the driving force $f_c$. In a small time interval $\delta t$, the interface at the kink position advances $\delta h = f_c \delta t$ in the vertical direction and 
$\delta x \approx \delta h \xi/2\pi$ in the horizontal direction. 
Therefore $c\approx f_c \xi/2\pi$.} and then each of them contributes equally to the velocity, so the velocity is proportional to the number $N$ of kink/anti-kink pairs present in the system.
The equilibrium value of $N$ is obtained by balancing the creation of kinks ($\sim R$) to its annihilation rate, which (like a chemical reaction between two species) is proportional to $N^2$~\cite{habib2000}. Namely
\begin{equation}
\frac{dN}{dt}\sim R -N^2
\end{equation}
By requiring equilibrium ($dN/dt=0$) we obtain 
\begin{equation}
v\sim N\sim R^{1/2}.
\label{eq:vproptoN}
\end{equation}
Therefore, 
\begin{equation}
v_{d=1}(T,\varepsilon) =T^{3/10} \mathcal G(\varepsilon^{5/4}/T) 
\label{v1d}
\end{equation}
where 
$\mathcal G\equiv \sqrt \mathcal F$. 
Alternatively we can also write
\begin{equation}
v_{d=1}(T,\varepsilon) =\varepsilon^{3/8} \widetilde{\mathcal G}(\varepsilon^{5/4}/T) 
\label{v1db}
\end{equation}
This expression is consistent with the form found in Ref.\cite{Buttiker1981}, which  
can be written (in our notation and units, and up to pre-exponential numerical constants) as
\begin{equation}
v_{d=1}(T,\varepsilon)\sim \frac{\varepsilon^{11/16}}{T^{1/4}}\exp{\left(-\frac{24}{5}\frac{(2\varepsilon)^{5/4}}{T}\right)}
\label{Bl}
\end{equation}

\begin{figure}
\includegraphics[width=8cm,clip=true]{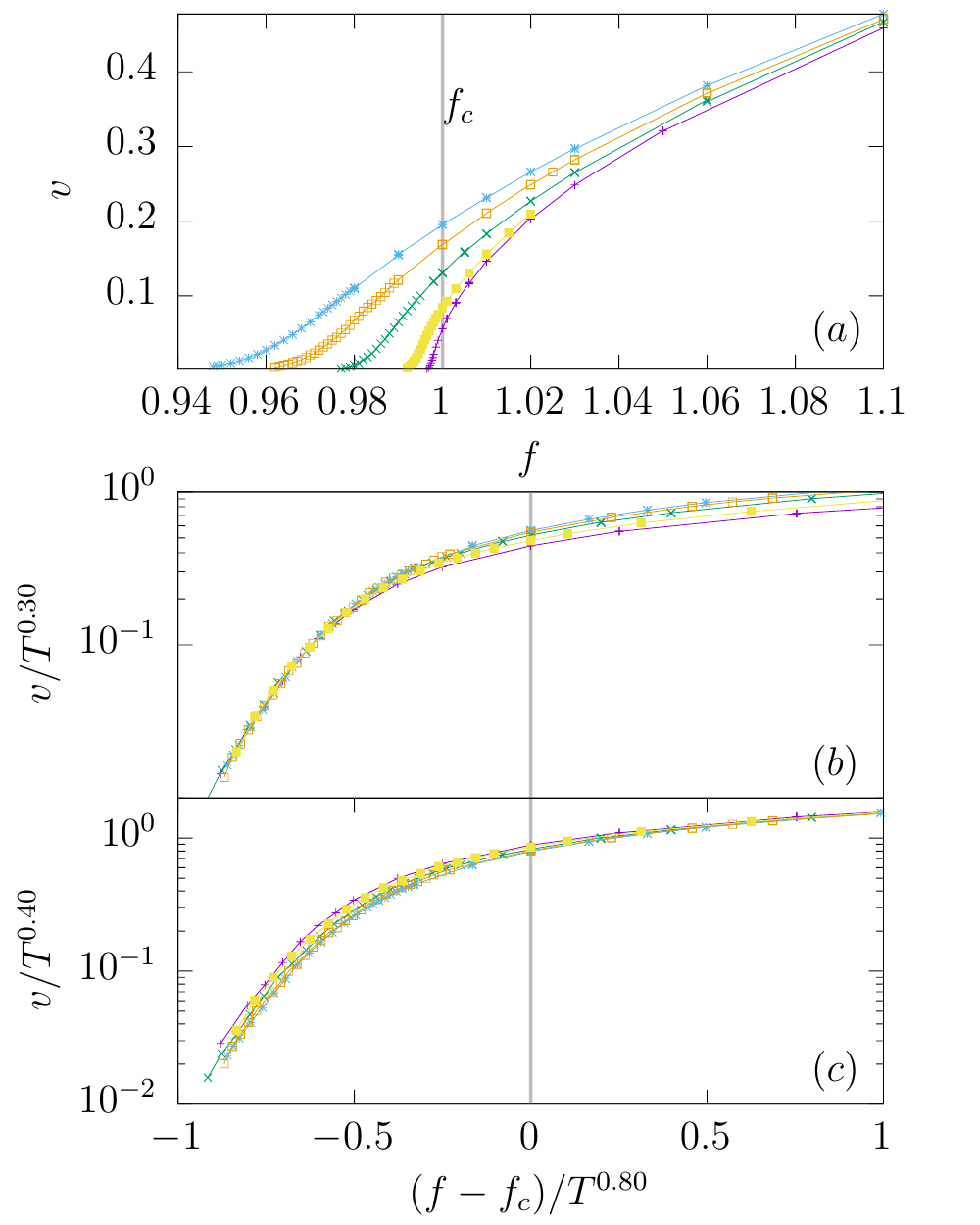}
\caption{Velocity force characteristics around the depinning 
threshold of an elastic string in the washboard potential,
at different temperatures $T=0.003,0.02,0.03,0.01,0.001$
(a) Whole range around $f_c=1$. Best scaling collapses 
just below the threshold $f\lesssim f_c$ (b) 
and just above the threshold $f\gtrsim f_c$ (c).
The exponents are clearly different above and 
below threshold, in contrast with the one particle case.
}
\label{vf1d}
\end{figure}

In Fig. \ref{vf1d} we test the scaling in Eq. (\ref{v1db}) numerically,
by integrating Eq. (\ref{eq0}) for $d=1$ 
using finite differencing and the stochastic Euler method on $L$ 
elastically coupled particles.
As shown in Fig. \ref{vf1d}(b) there is an excelent agreement 
below the depinning threshold, i.e., for $\varepsilon>0$. 
One remarkable thing about the 
scaling of Eq. (\ref{v1d}) is that it is {\it not} 
compatible with the well known behavior of $v$ for negative $\varepsilon$ 
(i.e. $f>1$) and $T=0$. In fact, if a $T$-independent limit is going to be extracted from Eq. (\ref{v1d}) this should be 
$v\sim |\varepsilon|^{3/8}$, 
that does not coincide with the known exact result 
$v(T=0, \varepsilon<0)\sim |\varepsilon|^{1/2}$.
This incompatibility can be appreciated in the $\varepsilon<0$ (i.e., $f>f_c$) part of Fig. \ref{vf1d}(b), where clearly the curves do not collapse. 
The good collapse for $f>f_c$ is obtained
rescaling $\varepsilon$ with 
the same energy exponent, namely 
$\varepsilon/T^{1/\alpha}= (f-f_c)/T^{1/\alpha}$ 
with $\alpha=(6-d)/4=5/4$ {\cite{Buttiker1981}}, 
but using $v/T^{2/5}$ in the vertical axis, in order to obtain $\beta=1/2$ (Fig. \ref{vf1d}(c)).
The conclusion is that a unique thermal rounding scaling is not valid in this problem for $d$ equal to (or larger than) one. 
In particular, if we try to define a single thermal rounding exponent, we should choose
$\psi=3/10$ from the $f<f_c$ part of the scaling, but $\psi=2/5$ from the $f>f_c$ part. We will see below
how this incompatibility manifests in the true form of $v(T)$  at $f=f_c$ having a non-trivial logarithmic correction.

\section{Thermal rounding of the depinning transition}

The results in Sec.~\ref{sec:activated}
clearly show that a unique global scaling of the form 
\begin{equation}
v=T^\psi \mathcal F(\varepsilon ^\alpha/T)
\label{naivescaling}
\end{equation}
is valid only for the simplest case of a single particle, but does not apply
to interfaces in finite dimensions.

For an extended interface the form of the activated dynamics scaling below $f_c$ ($\varepsilon >0$) 
cannot be extrapolated to the $\varepsilon<0$ region.  
Moreover, the form of the velocity 
as a function of temperature predicted by 
Eq.(\ref{naivescaling}) when $\varepsilon=0$, namely
\begin{equation}
v\sim T^\psi
\label{eq:naivethermalrounding}
\end{equation}
is not accurately satisfied, as we will show below.
It turns out that this scaling has important logarithmic 
corrections that we will now address.

We will make a detailed analysis of the dynamics of the system right at the critical force $f_c$ (i.e. $\varepsilon=0$).
Thus the model to be studied is that of Eq. (\ref{eq0}) 
for a sinusoidal pinning potential at the critical force, namely
\begin{equation}
\frac{\partial  h(r,t)}{\partial t}=-\cos(h)+1 +\nabla ^2 h +\sqrt T \eta(t,r),
\label{eq:sinegordon}
\end{equation}
as a function of temperature, in the $T \ll 1$ limit, 
where critical scaling functions and exponents are expected.

\begin{figure}
\includegraphics[width=8cm,clip=true]{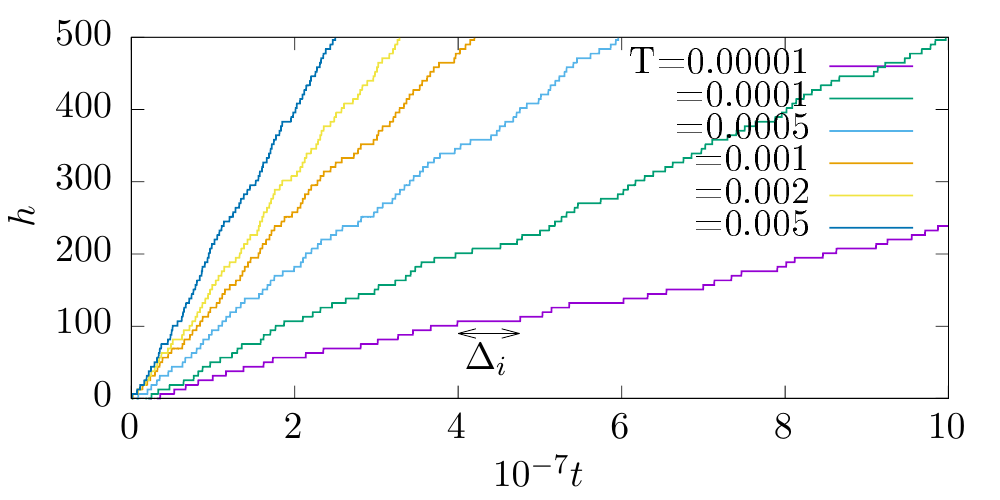}
\caption{Position-time plot for a single particle at the critical force, at different temperatures, as indicated. The motion proceeds by a sequence of jumps of size $2\pi$ separated by stochastic waiting times $\Delta_i$.
}
\label{1particula}
\end{figure}

\begin{figure}
\includegraphics[width=8cm,clip=true]{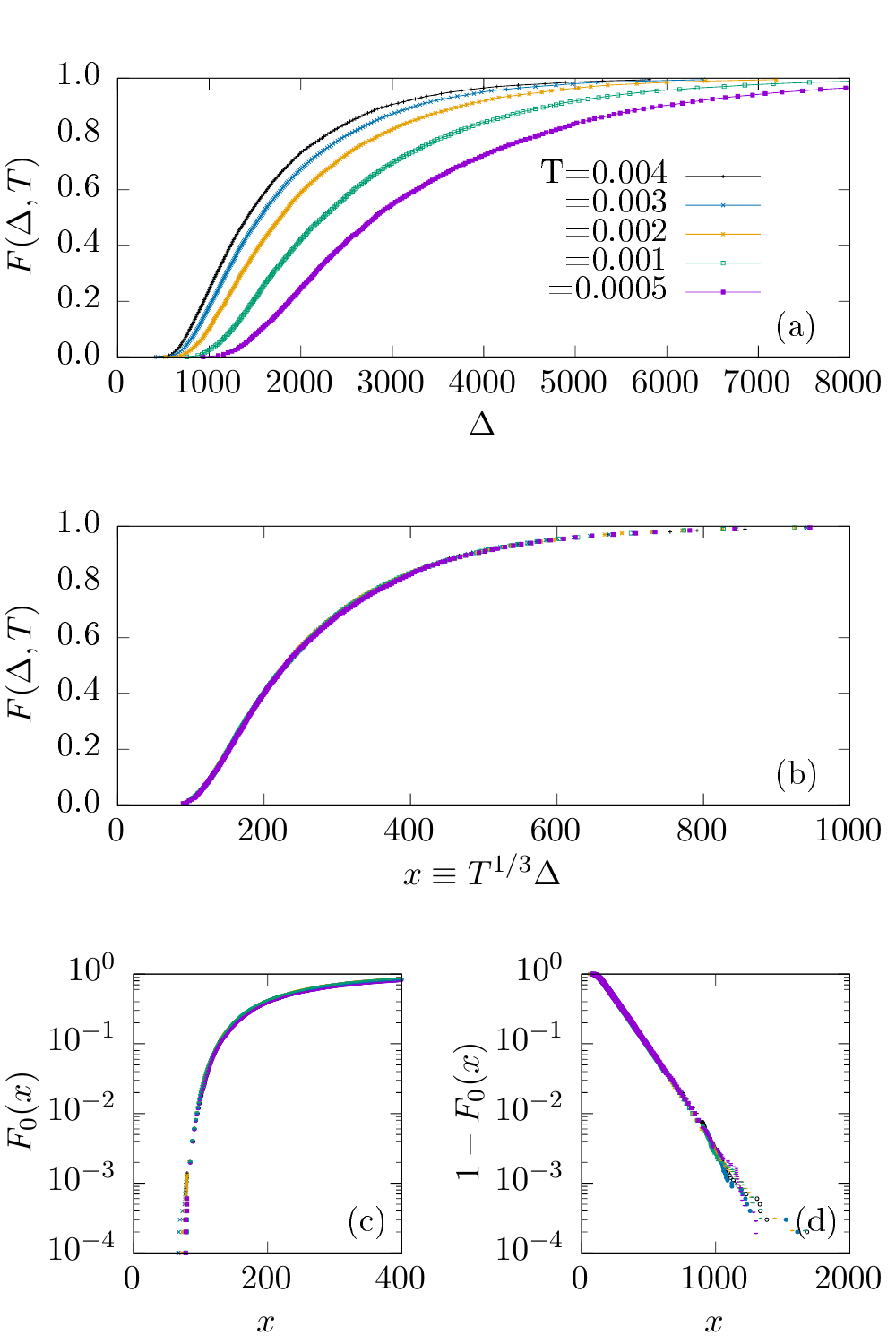}
\caption{
(a) Cumulative distribution 
[$F(\Delta,T)\equiv \text{Prob}(\Delta_i<\Delta)$] of the 
stochastic waiting times $\Delta_i$ (see Fig. \ref{1particula}) between successive jumps 
of a single particle ($d=0$) in a tilted washboard potential, 
exactly at $f=f_c=1$, for different temperatures.
(b) The distribution can be collapsed onto a single curve 
by plotting it as a function of $T^{1/3} \Delta$. 
The left (c) and right (d) tails of the cumulative distribution
highlight, respectively,  the slower-than-exponential growth of the probability 
for short waiting times and its exponential decay for large 
waiting times.
}
\label{PdeD1part}
\end{figure}

\subsection{Single Particle}
To serve as a reference we start with the analysis of the 
single particle case, that is, solving 
\begin{equation}
\frac{dh}{dt}=-\cos(h)+1 +\sqrt T \eta(t), 
\end{equation}
with $\langle \eta(t)\eta(t')\rangle=2\delta(t-t')$.
Fig. \ref{1particula} displays the numerically obtained evolution of $h(t)$ for different temperatures. We clearly see that the dynamics proceeds through abrupt jumps between successive ``bottlenecks'' positions that occur when $h$ is a multiple 
of $2\pi$, at which the particle spends most of the time. These are the points at which the deterministic force on the particle vanishes.
The average velocity as a function of temperature follows the prediction of Eq. (\ref{vd0}) at $\varepsilon=0$, namely $v\sim T^{1/3}$. However we emphasize that this scaling applies not only to the average velocity (which is related to the average waiting time at the bottlenecks) but also to the whole distribution of time intervals  spent at the bottleneck positions. This is shown in Fig. \ref{PdeD1part}(a) where the cumulative 
probability distribution $F(\Delta,T)$ of the time intervals $\Delta$ spent at each bottleneck is calculated 
for different small temperatures
\footnote{
As a matter of definition, the values of $\Delta$ are calculated as the time interval between the time $h$ crosses the value $2\pi (n-1/2)$ and the time in which it crosses 
the value $2\pi (n+1/2)$
}. 
As shown in Fig. \ref{PdeD1part}(b) the results adjust perfectly to the scaling law 
\begin{equation}
F(\Delta,T)=f(\Delta /T^{1/3})
\label{1}
\end{equation}
Therefore, for the average particle velocity $v\sim \overline{\Delta^{-1}}$ we get
$v \sim T^{1/3}$, 
as it has been observed with high accuracy (see for instance 
Ref. \onlinecite{purrello2017}, and Fig.\ref{vd0}). 
This simply confirms that the particle accurately obeys the thermal rounding 
scaling of Eq. (\ref{eq:naivethermalrounding}).

As shown in Fig. \ref{PdeD1part}(c) and (d), 
$f(\Delta)$ displays an exponential decay at large $\Delta$ and a sort of ``pseudogap'' at small $\Delta$, where $f(T^{1/3}\Delta)$ 
is almost zero. 
The existence of this ``minimum time'' for a jump 
\footnote{The strong decay of $f(x)$ we observe by 
decreasing $x$ and small temperatures 
makes it that in practice, it can be considered as a hard gap.}
will play an important role in the analysis of the movement of the one-dimensional string, that we consider in the following. 

\subsection{Elastic String}

\subsubsection{Kink/anti-kink dynamics at $f=f_c$}
\label{sec:kinkantikinkfc}
\begin{figure}
\includegraphics[width=8cm,clip=true]{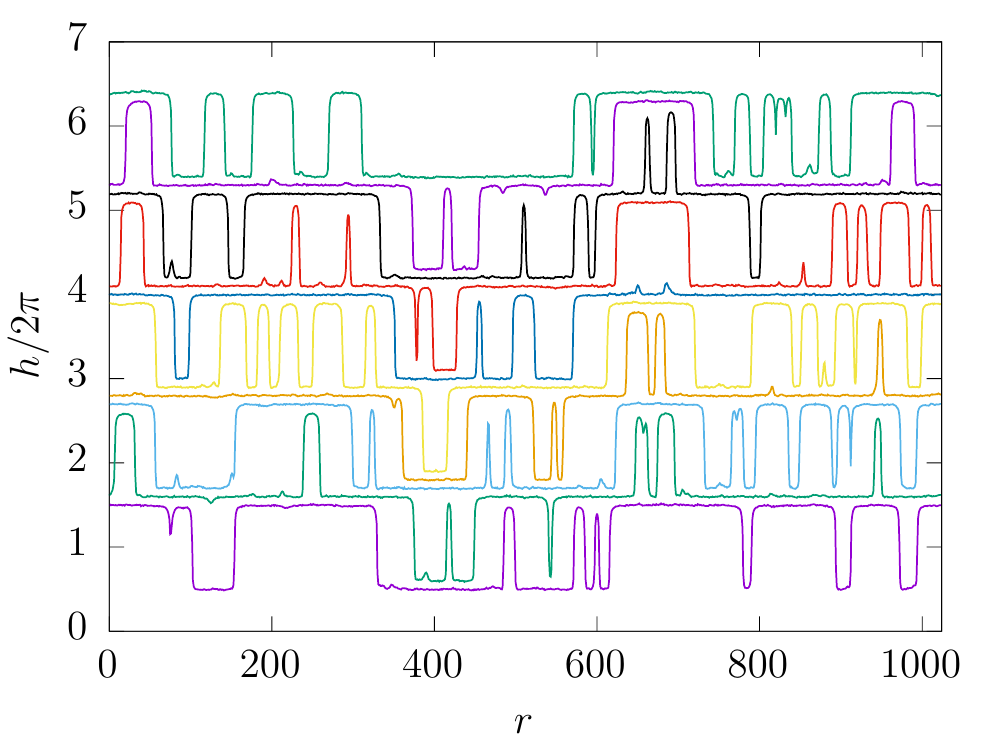}
\caption{
A sequence of numerically generated configurations 
$h(r,t)$ for an elastic string of size $L=1024$, 
$T=5\times 10^{-5}$, exactly at the critical force $f_c=1$.
Consecutive configurations at increasing times (from bottom to top) 
have been differently colored and slightly shifted 
vertically by $\sim 0.1$ for clarity.
}
\label{configsh}
\end{figure}

\begin{figure}
\includegraphics[width=9cm,clip=true]{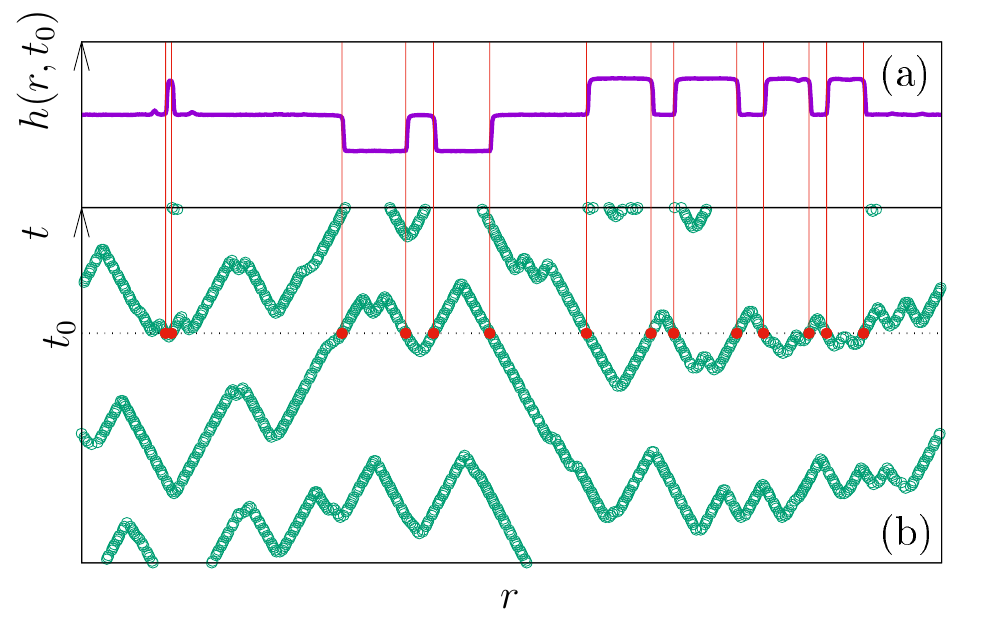}
\caption{
(a) Snapshot of a configuration $h(r,t_0)$ 
of the interface at $t=t_0$, exactly at the critical force 
$f_c=1$ and $T=10^{-5}$, generated from Eq. (\ref{eq:sinegordon})
for $L=1024$. (b) Kink trajectories in space-time. 
Red dots correspond to kink positions at $t=t_0$.  
Dashed vertical lines show their correspondence 
with $h(r,t_0)$ kinks.
}
\label{configshkinks}
\end{figure}

\begin{figure}
\includegraphics[width=8cm,clip=true]{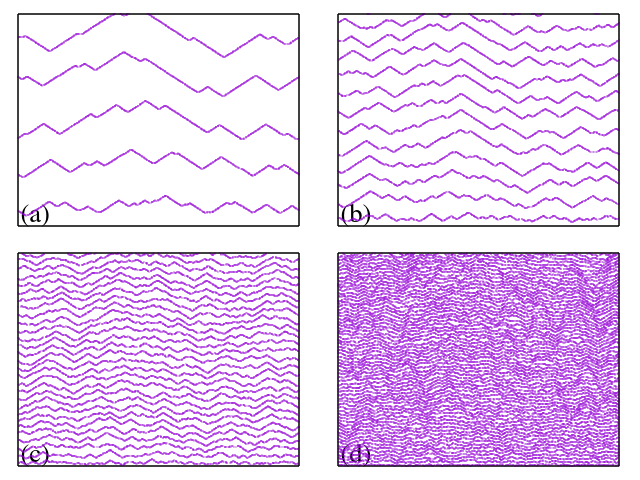}
\caption{
Numerically generated kink trajectories exactly at the critical force for 
different temperatures, $T=10^{-6}$ (a), $T=10^{-5}$ (b), $T=10^{-4}$ (c), 
and $T=10^{-3}$ (d), for a string of size $L=1024$. 
The vertical and horizontal directions are 
time and position respectively, and have the same extent in the four panels. 
The time-gaps between the zig-zagging contours formed 
by many kink and antikink trajectories arise naturally from the dynamics.
}
\label{kinktrajectoriesfull}
\end{figure}

To gain insight in the form in which a one-dimensional elastic string moves at $f=f_c$,
we solve numerically Eq. (\ref{eq:sinegordon}) for $d=1$.
 In Fig. \ref{configsh} we display a few snapshots of the configuration of the system in a well equilibrated state, at slightly increasing times. We see a characteristic structure in which pieces of the interface are located at positions corresponding to the bottlenecks of the potential. 
For convenience we will number successive bottlenecks with an integer index $\nu$, such that the interface stays at $h=2\pi \nu$.
Different pieces of the interface are connected through ``kinks'' in which the interface passes from $\nu$ to $\nu\pm 1$, as again shown in Fig. \ref{configshkinks}(a). 
It is important to realize that the kinks move in a very deterministic and predictable way. In fact, 
as a piece of interface at position $\nu$ has a potential energy per site of $ -2\pi \nu$, a kink connecting $\nu$ and $\nu+1$ decreases its energy by moving in the direction 
that increases the extent at $\nu+1$ an decreases that at $\nu$.
This produces that all kinks in the system move at a constant velocity 
$c \approx f_c\xi/2\pi${\cite{Buttiker1981}}, where $\xi$ is the kink width. For our 
numerical setup (Eq. (\ref{eq:sinegordon})) we find 
$c \approx 0.24$, in consistence with $f_c=1$ and the observed  
$\xi \approx O(1)$ (see Fig. \ref{configsh}),
always in the direction of producing a net advance of the interface.

Fig. \ref{configshkinks}(b) is an alternative and comprehensive view of kink movement in the system. 
It is a space-time plot of all kinks or anti-kinks trajectories in the system.
As kinks move always at the velocity $\pm c$, their trajectories are seen as straight lines  in  Fig. \ref{configshkinks}(b). When kink and anti-kinks collide, they annihilate at the ``$\Lambda$''-shaped 
points. In addition, kink/anti-kink pairs nucleate at the ``V''-shaped points. 
Note that Fig. \ref{configshkinks}(b) can be described as a ``contour plot'' of the funcion $h(x,t)$
, each contour (characterized by an increasing integer number $\nu$) indicating the time at which the interface first reached the  height $h=2\pi\nu$.
The space-time plot of Fig. \ref{configshkinks}(b) is a full picture of the dynamics of the string in spatial scales larger than the typical size of the kinks.
Fig. \ref{configshkinks}(b) hence reveals the sparse and localized 
activity of the interface.

In Fig. \ref{kinktrajectoriesfull} we can qualitatively appreciate the kink dynamics at 
$f=f_c$ for different temperatures in the steady-state. The four panels correspond to four different increasing temperatures.
The space and time extent of the four panels is the same.
We observe in particular that the slope of the straight segments (corresponding to kink propagation) 
has the same value $c^{-1}$ for all temperatures.
As described above the space-time 
segments describing kink trajectories form a well defined 
sequence of activity contours
that percolate in space but are
separated by distinguishable time-gaps 
(i.e., different lines do not get close vertically in practice). 
It is worth stressing however that, in spite of these time gaps, 
the one dimensional interface at the steady-state 
is actually {\it never} completely trapped 
in a metastable state: for given time $t$ a large enough interface 
always has pairs of kinks evolving quasi-deterministically. 
In other words, a line 
with $t=\text{const}$ in space-time always cuts the trajectory of 
some kinks in a thermodynamic system at any finite temperature. 
Another interesting property that can be appreciated in 
Fig. \ref{kinktrajectoriesfull} is 
that increasing temperature increases both the space-time density of annihilation 
and creation events, and decreases the time gaps, strongly suggesting 
a space-time-temperature scaling relation, 
that we will discuss now in detail.   

\begin{figure}
\includegraphics[width=8cm,clip=true]{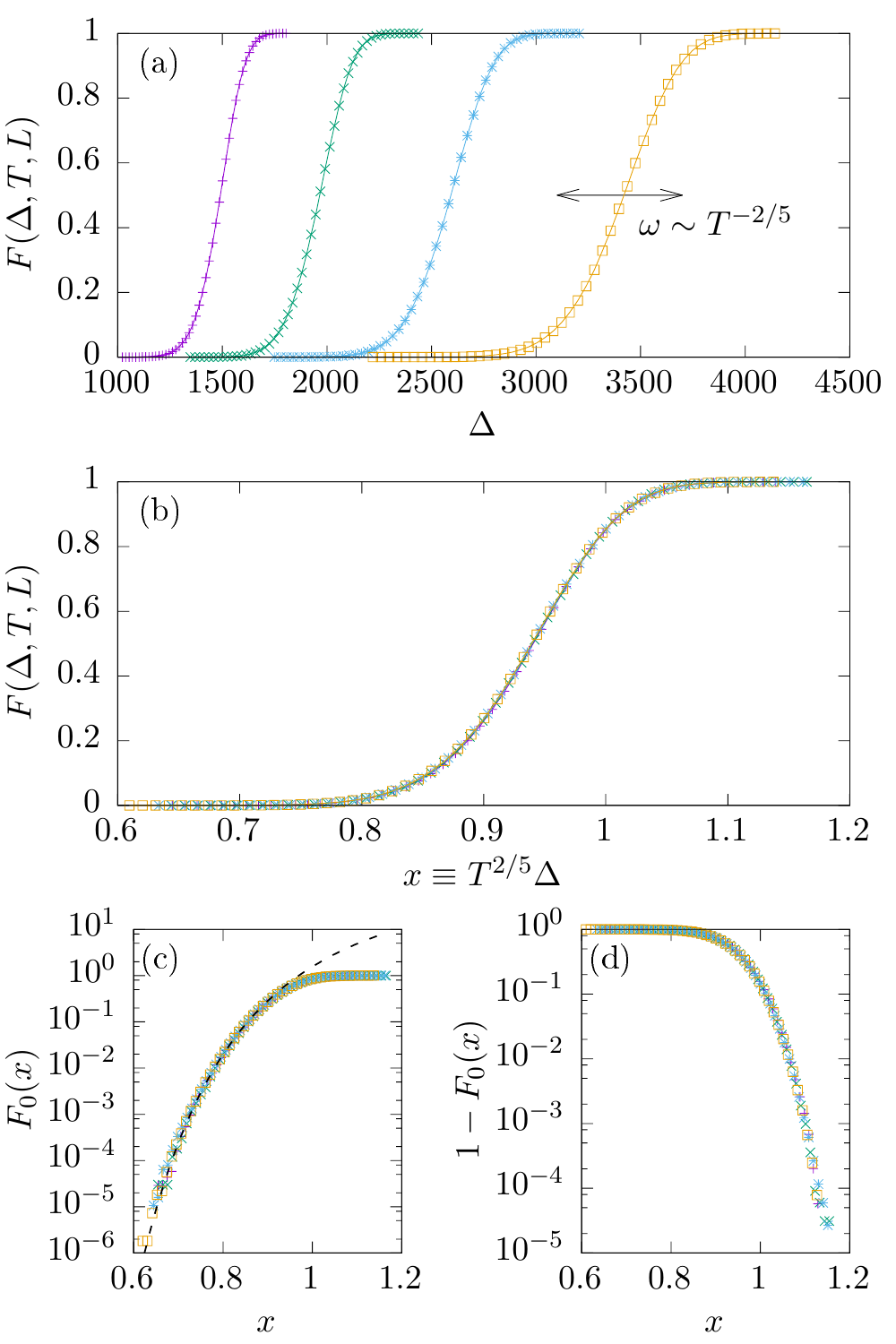}
\caption{
Waiting-time cumulative distributions for the activation of the first 
pair of kinks in a flat ($d=1$) elastic string exactly at 
$f=f_c$, for different temperatures $T$ and string sizes $L$, 
such that $T L^{1/5} = \text{const}$.
(a) The rightmost curve corresponds to
$L=2^{16},T=1.25 \times 10^{-9}$, 
each following curve 
to the left doubles the temperature. 
(b) Master curve using the time-temperature-length scaling, 
$F(\Delta,T,L) \sim {\tilde F}(T^{2/(6-d)}\Delta, T^{1/(6-d)} L)
\equiv F_0(T^{2/(6-d)}\Delta)$ in the case $d=1$.
(c) The dashed line
$\sim \exp[-4.58/x^3]$ is an empiric fit of the left 
part of $F_0(x)$. 
(d) Detail of the right tail of $F_0(x)$, 
to be compared with the single particle case in Fig.\ref{PdeD1part}.
\label{fig:FdeDelta}
}
\end{figure}

In Sec.\ref{sec:activated}, and working below the critical force, we considered 
the nucleation of kink/anti-kink pairs to occur at a rate $R$ that was simply a function of temperature. 
This led to the estimation that the velocity interface is $v\sim R^{1/2}$ 
(Eq. (\ref{eq:vproptoN})). 
This analysis was appropriate because in that case there was a finite energy barrier to be surmounted, and the dynamics of this activation is statistically a Poisson process: if an attempt to climb the barrier has failed, the next one has to start over, independently of how many previous attempts have been made. 
But right at $f_c$, the transition between successive bottleneck positions does not require the climbing of any energy barrier. The bottlenecks are characterized by a flat potential in which the deterministic force vanishes, and the transition time displays the typical time gap already seen in the single particle case (Fig. \ref{PdeD1part}). 
Therefore in this case the nucleation rate $R$ previously used is not a useful concept. Instead, it will be useful to consider (as in the 0-dimensional case) the function $F$, that measures the time in which a kink/anti-kink pair is first observed starting with an originally  flat interface at $f_c$.

We consider a system of size $L$ (with periodic boundary conditions), and start at $t=0$ with a flat configuration at $h=h_0<0$. This configuration moves deterministically towards the saddle located at $h=0$, and would remain there if $T=0$. However, thermal fluctuations produce the surpassing of the saddle, and we determine the time $\Delta$ at which the first point of the interface is detected at some positive $h_1$, indicating a kink/anti-kink pair has been created. The value of $\Delta$ becomes large as $T\to 0$, and therefore the precise values of $h_0$ and  $h_1$ are not important. We choose $h_0=-1$, $h_1=1$. The numerical procedure is repeated many times to collect statistics of $\Delta$ values.
Fig. \ref{fig:FdeDelta}(a) shows the cumulative distribution function $F$ of the first nucleation time $\Delta$, 
for different temperatures $T$ and system sizes $L$. 
The form of $F(\Delta,T,L)$ can be simplified using the general scaling theory of Section III, that can be applied to the present calculation without change, simply considering $\varepsilon\equiv 0$. Since $F$ is a dimensionless function, it must remain the same when its arguments are changed according to the scaling in Section III.
Therefore, we obtain:
\begin{equation}
F(\Delta,T,L)={\tilde F}(T^{2/(6-d)}\Delta ,T^{1/(6-d)}L).
\label{eq:Fscaling}
\end{equation}
Since $d=1$, and we are fixing $L T^{1/5}$ in the simulations, for simplicity we will omit the $L$ dependence everywhere
and simply write 
$F(\Delta,T,L)= {F}_0(x)$ with $x \equiv T^{2/5}\Delta$
and ${F}_0(x)$ the master curve. 
In panel (b) of Fig. \ref{fig:FdeDelta} we show the excellent collapse obtained for the different curves in (a) obtained using this scaling.

It is interesting to compare the tails of $F_0(x)$ for the string 
and the particle. In the single particle case $F_0(x)$ is exponential 
for large $x$, while the elastic string displays a clear faster-than-exponential decay at large $x$. On the other hand for 
small $x$, both cumulative distributions display a slower than exponential 
growth.

The results in Fig. \ref{fig:FdeDelta} clearly display the ``gap" effect in the nucleation time (also observed for the single particle case in Fig.\ref{PdeD1part}), pointing also to the fact that this nucleation cannot be considered anymore (as it was in the activated regime)
a Poisson process.
Thus the probability to nucleate a kink/anti-kink pair in a piece of interface at height $h=2\pi \nu$ depends on how much time the surface has stayed at $2\pi \nu$ already.
To get an idea of this phenomenon and its importance, it is 
worth looking again in Fig. \ref{kinktrajectoriesfull} 
to the ubiquitous time-gaps appearing in sequences of many contours
for different temperatures. 
The rather well defined values of the vertical gaps between different contours in the plot in Fig. \ref{kinktrajectoriesfull} is a consequence of the fact that nucleation time cannot be arbitrarily small, as seen also in 
Fig.~\ref{fig:FdeDelta}. The numerical results for the function $F$ just presented will be useful in the next Section to calculate the interface velocity at $f=f_c$.

\subsubsection{$v(T)$ curve at $f=f_c$}

Armed with the qualitative understanding of the dynamics we gained in the previous 
section (Sec.\ref{sec:kinkantikinkfc}), we can address quantitatively the expected form of the temperature dependence of the velocity $v$ right ar $f_c$, namely the thermal rounding law. Referring to the plots in Fig. \ref{kinktrajectoriesfull}, the value of $v$ is nothing more than $2\pi$ divided by the average temporal separation
between successive contours, that we call $\Delta_0$. We must estimate $\Delta_0$ in order to calculate $v$ as $v=2\pi /\Delta_0$.

\begin{figure}
\includegraphics[width=8cm,clip=true]{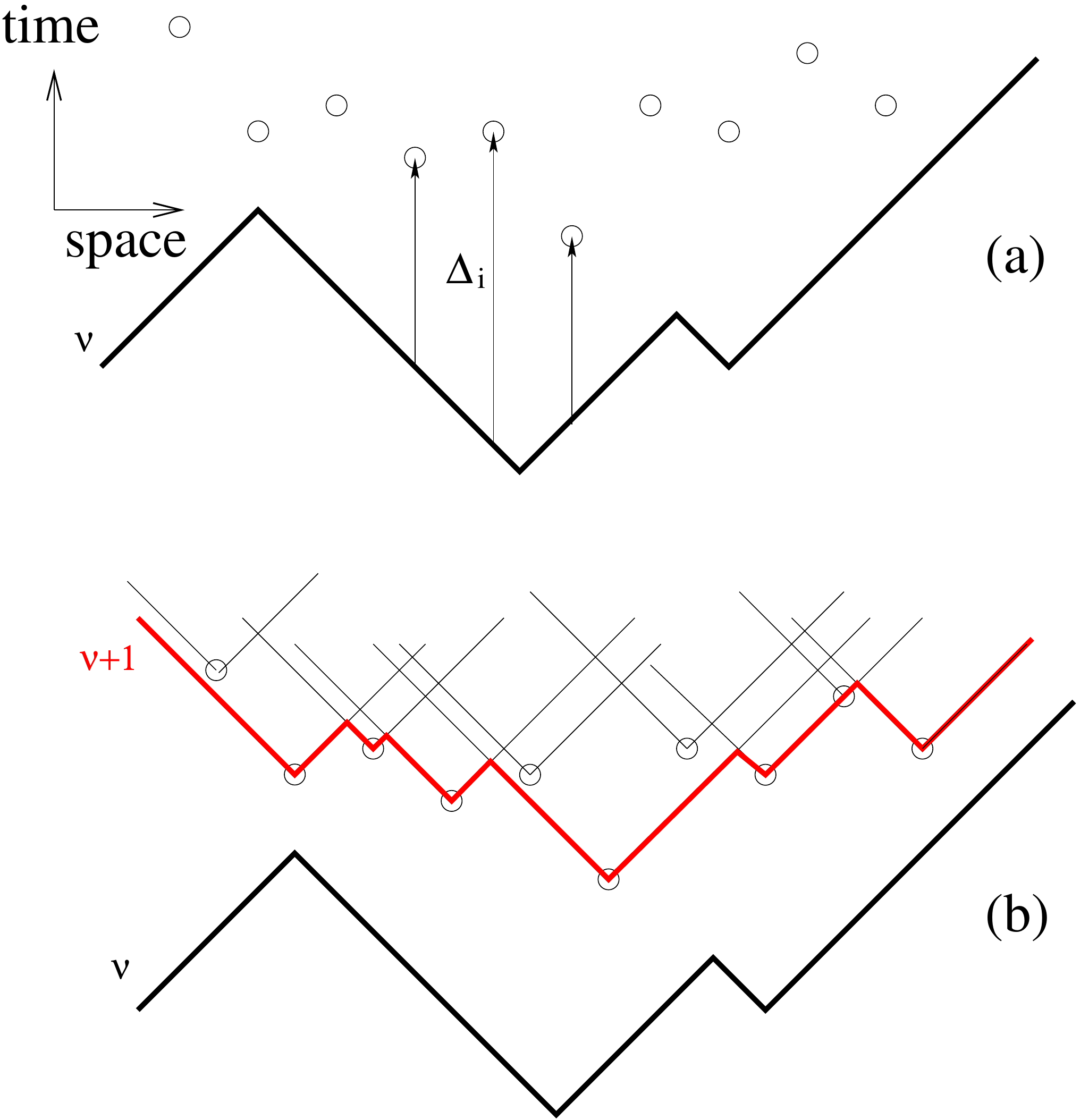}
\caption{(a) Thick line: Sketch of contour $\nu$ of the space-time configuration of an interface (see Fig. \ref{kinktrajectoriesfull}). In principle, at each spatial position, the next kink/anti-kink is nucleated at the time indicated by the small circles. (b) The actual $\nu+1$ contour is constructed by kinks and anti-kinks emanating from the nucleation points. Only the earliest kink/anti-kink actually contribute to the contour $\nu+1$ (shown in red). Notice that the average time position of contour $\nu+1$ is smaller than the average time position of all circles.
}
\label{sketch}
\end{figure}

One may naively expect that $\Delta_0$ is simply given by the average value of the nucleation time $\overline \Delta$ that can be extracted from the data in Fig. \ref{fig:FdeDelta}. However, this is not quite so. Kink/anti-kink pairs nucleated with particularly small values of $\Delta$ (at the left of the distribution in Fig.  \ref{fig:FdeDelta}) will have a stronger influence, and produce that $\Delta_0<\overline \Delta$.
In order to understand correctly this fact and its importance, let us consider the sketch in Fig.~\ref{sketch}. In panel (a) we depict with the thick black line one of the contours (with label $\nu$) already shown in Fig. \ref{kinktrajectoriesfull}.
This contour represents, for all spatial positions, the time at which the interface reaches the height $2\pi\nu$. Suppose that each horizontal position in Fig. \ref{sketch} represents a portion of the interface to which the analysis in Fig. \ref{fig:FdeDelta} can be applied. This means that for each horizontal position we can draw a point from Fig. \ref{fig:FdeDelta} and plot a {\em nominal time} at which a kink/anti-kink pair would be nucleated at that position. This is represented in Fig. \ref{sketch}(a) as the open circles. The average temporal distance between contour $\nu$ and all open circles is simply the value $\overline\Delta$ extracted from the  Fig. \ref{fig:FdeDelta}. However, each nucleation point generates a kink/anti-kink pair that propagates in the system as indicated in Fig. \ref{sketch}(b). The actual $\nu+1$ contour is the 
lower envelope of all these kink/anti-kink pairs. It is apparent from Fig. \ref{sketch}
that the average time separation $\Delta_0$ between contours $\nu$ and  $\nu+1$ is smaller than $\overline {\Delta}$.

In order to calculate $\Delta_0$ explicitly we notice that the contour $\nu+1$ is composed by kinks and anti-kinks originated in the lowest values of the nucleation times at all sites, or in other words the lowest circles in Fig. \ref{sketch}. If $M$ is the average number of sites affected by a single kink/anti-kink pair, the typical value of $\Delta_0$ corresponds to the typical value of the minimum of $M$ variables
$\Delta_i$ with cumulative distribution $F(\Delta,T)$ (the function plotted in Fig. \ref{fig:FdeDelta}). This minimum is roughly given by the condition 
\begin{equation}
MF(\Delta_0,T)=1,
\label{mp}
\end{equation}
and the velocity of the interface will be given by $v=2\pi/\Delta_0$. The main temperature dependence of $v$ comes from the temperature dependence of  $F(\Delta_0,T)$. The velocity dependence of $M$ will account for a logarithmic correction, as we will now see. From the numerical results in Fig. \ref{fig:FdeDelta}(c), $F(\Delta_0,T)$ can be very well approximated (particularly when $F\ll 1$) as
\begin{equation}
F(\Delta_0,T)\simeq \exp(-C/(T^{2/5}\Delta_0)^3)
\label{p1d}
\end{equation}
with $C\simeq 4.58$ a numerical constant.
Note that the form of the combination $T^{2/5}\Delta$ comes already from the scaling of time and temperature, in the analysis of Section \ref{sec:activated} applied to the $d=1$ case. The third power instead, is just a rough numerical fitting (see dashed-line in Fig.\ref{fig:FdeDelta}(c)).
The dependence of $M$ on temperature is roughly given by the following argument:  a kink/anti-kink starting at one of the lowest circles in Fig. \ref{sketch} will be part of the $\nu+1$ contour for a number of sites $M$, such that $M/c\sim \omega$, where $\omega$ is the width of the $F(\Delta,T)$ function. According to  
Eq. (\ref{p1d}) $\omega$ scales with temperature as $\omega \sim T^{-2/5}$. Putting the pieces together, and since $c$ is just a constant, this gives simply
\begin{equation}
M\sim  T^{-2/5}
\label{m}
\end{equation}
Using now Eqs. (\ref{p1d}) and (\ref{m}) in Eq. (\ref{mp}) we finally obtain

\begin{equation}
v=c_1 T^{\frac 25}[-\log(c_2T)]^{\frac13}
\label{v}
\end{equation}
where we have dig into $c_1$ and $c_2$ all unknown constants of the analysis.
In general dimension $d$, the kink-antikink 
pair of the $d=1$ case is replaced by a $(d-1)$-dimensional 
domain wall describing a droplet boundary, 
allowing $d$-dimensional 
patches to advance from one position to the following 
in an isotropic way.
The mechanism just described of nucleation, expansion 
and coalescence of defects qualitatively applies in 
general dimension 
and the expected form of $v$ at $f_c$ is
\begin{equation}
v=c_1 T^{\frac 2 {(6-d)}}[-\log(c_2T)]^{\delta}
\label{v2}
\end{equation}
where the exponent $\delta$ of the logarithmic correction is 0 in $d=0$, and $1/3$ in $d=1$. For $d>1$ we expect it to be different from zero, but we have not attempted a precise determination.

The result we have obtained for the dependence of $v(T)$ shows a main power law dependence but also an important logarithmic correction that can have an important effect on experimentally observed values. Qualitatively, the origin of the two parts can be traced back to the particular dynamics of the problem. The $T^{2/(6-d)}$ factor in the velocity comes from the average transition time between bottleneck configurations at which the interface spends most of the time. The logarithmic factor is a consequence of the linear-in-time increase of the extent of the interface at position $2\pi \nu$ before nucleating the defect that will allow the transition to the $2\pi(\nu+1)$ position (see appendix B for a more formal derivation of the necessity of such a logarithmic correction, independently of the details of the dynamics).

\begin{figure}
\includegraphics[width=8cm,clip=true]{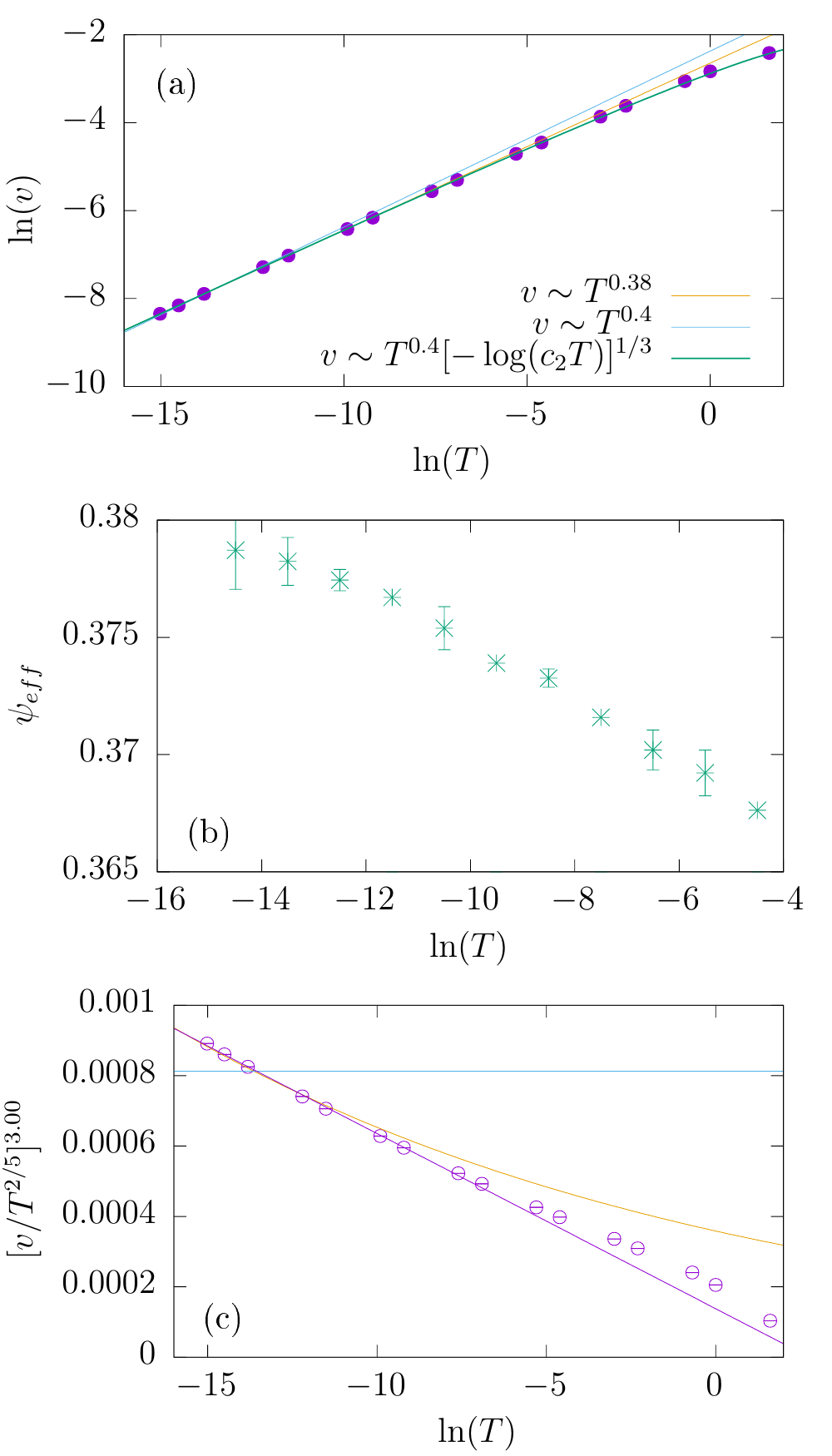}
\caption{
(a) Velocity as a function of temperature ($T \in [10^{-7},10]$) at $f=f_c$, 
for an elastic string of size $L=8388608$. 
The yellow line is a naive pure power-law fitting of the form $v\simeq T^{\psi_0}$,
$\psi_0\simeq 0.38$.  
The green line is a fit from our Eq. (\ref{v}), consisting in a power law 
$T^{2/(6-d)}=T^{0.4}$ multiplied by a logarithmic correction. 
The blue line, with slope $2/(6-d)=0.4$, is used only for reference. (b) Effective 
power-law exponent 
$\psi_{eff}\approx d\log v/d \log T$, using the method of 
consecutive slopes (see text). 
(c) The same results as in (a) but plotted as $[v/T^{0.4}]^3$ vs $\log T$, 
highlighting the logarithmic corrections. The accuracy of Eq. (\ref{v})
is very good, except for large temperatures where 
some non-considered effects enter into play.
}
\label{vdet}
\end{figure}

We now check the form of $v(T)$ from Eq. (\ref{v}) against numerical simulations.
The results span seven orders of magnitude in temperature ($10^{-7} \leq T \leq 1$) 
in a large enough system ($L=2^{23}$) such that $T^{1/5}L \gg 1$ so 
to avoid finite-size effects (also implying a large number of 
evolving kinks at any instant).
Fig. \ref{vdet}(a) shows the results of a simulation in the full model at $f=f_c$. If we were trying to fit a power law, we would probably fit an exponent $\simeq 0.38$ (yellow line) at least in the left part of the figure. Yet, our proposed behavior (Eq. (\ref{v}))
produces a more satisfactory and consistent result. By fitting appropriately $c_1$ and $c_2$ we find the green curve, that fits the data in a much broader range of temperatures. 
This is even clearer in panel (b) 
where an effective 
power-law exponent as a function of temperature is obtained
$\psi_{\tt eff}\approx d\log v/d \log T$, using the method of 
consecutive slopes, and fitting pure power-law in windows 
of size $[T-\Delta T,T+\Delta T]$ with $\log((T+\Delta T)/(T-\Delta T))=3$.
This effective exponent shows a dependence compatible with very slow convergence to 0.4 when $T\to 0$, as Eq. (\ref{v}) implies.
Also, in panel (c) the data are plotted in such a way that they must follow a straight line if Eq. (\ref{v}) is followed.
We see in fact that they follow very well this behavior, 
except for large temperatures in which some effects not considered in our analysis enter into play (particularly when temperature becomes a sizeable fraction of the total amplitude of the corrugation potential and the system crossovers to the fast-flow regime where $v\sim f_c$).

\section{Conclusions}
The naive analogy of the depinning transition with 
standard phase transitions suggests that the average 
velocity of an extended elastic manifold exactly at the threshold should scale
as $v\sim T^\psi$ for small temperatures $T$, 
with $\psi$ the thermal rounding exponent. 
Pioneer arguments testing this idea, 
and yielding the first non-trivial predictions for $\psi$, 
were first given in the context of charge density wave models with 
quenched disorder
~\cite{Fisher1985,Middleton1992a,Middleton1992b}
and 
later proposed for models of disordered elastic interfaces. 
In particular, they led to the relation $\psi=\beta/\alpha$, with $\beta$ 
the zero temperature velocity exponent 
($v(f,T=0)\sim (f-f_c)^\beta$) and $\alpha$ the barrier 
exponent describing how barriers $U$ for nucleation of 
forward moving modes vanish approaching the depinning threshold from 
below ($U(f)\sim (f_c-f)^\alpha$).
In spite of several subsequent 
analytical \cite{chauve2000,Nattermann2001,Muller2001}, numerical \cite{Chen1995,Roters1999,Bustingorry_2007,bustingorry2009thermal,
Bustingorry2012,Ferrero2013,purrello2017} 
(some of them with different 
predictions for $\psi$) and experimental 
\cite{Bustingorry2012,Gorchon2014,Pardo2017}
studies, a proper understanding of the thermal 
rounding of the depinning transition remains
elusive.

We have analyzed a simple version model of the depinning 
transition, namely the interface in a washboard potential, 
and found that right at the threshold 
the velocity follows by Eq.(\ref{v2}),
which contains an important logarithmic correction (when $d\ne 0$)
compared with the
pure power-law behavior.
We have shown that the logarithmic 
correction
in this 
model can be physically 
explained in terms of a competition between 
the droplet nucleations
(bounded by a kink-antikink pair in $d=1$ 
or a $d-1$ dimensional domain wall for $d>1$)
and the expanding deterministic motion 
they immediately trigger. In either case the 
later deterministic 
motion is hence not only responsible 
for displacing pieces of the interface one period 
further but also responsible for the 
deactivation of the nucleation 
in nearby sites (See Fig.\ref{sketch}). 
It is worth stressing that the left tail of the 
waiting time distribution for nucleation 
plays a fundamental role in producing logarithmic corrections.
At this respect we note that the characteristic space-time 
structure we observe 
at the depinning transition (see Fig \ref{kinktrajectoriesfull}) 
is clearly different from the one observed in the 
poly-nuclear growth model ~\cite{Goldenfeld_1984} 
(and other similar solid-on-solid growth models) 
where droplet nucleations are randomly 
sampled from a Poisson distribution 
before expanding them. As shown in 
Fig.\ref{fig:FdeDelta}, a Poisson 
distribution does not apply at all 
to the thermally assisted dynamics at the critical force. 
Interestingly, the exponent 
${2/(6-d)}$ in Eq.(\ref{v2}) still agrees 
exactly with the relation $\psi=\beta/\alpha$ 
(with $\beta=1/2$ and $\alpha=(6-d)/4$) proposed in 
Ref.~\onlinecite{Middleton1992b}. 
Furthermore, we have shown that the same prefactor exponent 
$\beta/\alpha$ actually holds for an infinite family of periodic 
potentials with different values of $\alpha$ and $\beta$
~(Section \ref{gralnormalform}). Our results hence predict 
that, in practice, the effective thermal rounding exponent will 
approach $\beta/\alpha$ slowly and from below 
in the limit of small temperatures.

Based on the above findings, we conjecture that 
the thermal rounding of the depinning transition 
in the more generic case of interfaces in disordered pinning 
landscapes for $d>0$ also displays logarithmic corrections, 
which may be written as 
\begin{equation}
v=c_1 T^{\beta/\alpha}[-\log(c_2T)]^{\delta}
\end{equation}
where $\delta \geq 0$ is a new exponent describing 
the left tail (or ``pseudogap'') of the waiting time distribution for nucleation of 
localized modes exactly at the critical force.
Such modes may be related to the marginally stable 
localized (at the Larkin length scale) soft modes 
found at the critical depinning configuration
~\cite{Cao2018}. In this scenario, the 
deterministic expansion of thermally 
nucleated droplets 
would be replaced by the analogous avalanche motion 
observed near the depinning threshold. 
Noteworthy, some interface 
models with disorder gave already evidence of 
logarithmic corrections~\cite{purrello2017}. 
Testing this conjecture more broadly may help to 
advance our understanding of the thermal rounding of 
the depinning transition of elastic manifolds.

\acknowledgments

We thank S. Bustingorry, E. Ferrero and V. Lecomte for stimulating discussions. We also acknowledge support from grants PICT2016-0069 (MinCyT) y UNCuyo2019-06/C578.

\appendix
\section{Generalization of the nucleation rate scaling and thermal rounding}
\label{gralnormalform}
Here we consider the motion of a flat interface segment 
near the depinning transition ($f=f_c-\varepsilon$) using a more general form 
for the bottleneck at $h=0$,
\begin{equation}
\frac{\partial h(r)}{\partial t}=h^\gamma -\varepsilon +\nabla ^2 h+\sqrt T \eta(t,r)
\label{eq1gamma}
\end{equation}
with $\gamma$ characterizing the normal 
form of the periodic force $-V'(h)\approx -h^\gamma$ around $h=0$ 
and all its periodic images.  
Using the same arguments leading to \ref{eq:scaling}
we now arrive to its generalization,
\begin{eqnarray}
\varepsilon &\to& \tilde \varepsilon\equiv k \varepsilon\\
h &\to& \tilde h \equiv k^{\frac{1}{\gamma}}h\\
t &\to& \tilde t \equiv   k^{\frac{1}{\gamma}-1} t\\
z &\to& \tilde z \equiv k^{\frac{1-\gamma}{2\gamma}} z\\
T &\to& \tilde T \equiv k^{2-\frac{(2+d)(\gamma-1)}{2\gamma}}  T
\end{eqnarray}
which reduces for $\gamma=2$ to Eq. (\ref{eq:scaling}).
Repeating the same steps than for $\gamma=2$ we obtain the 
generalized $\gamma$ dependent nucleation rate per unit volume
\begin{equation}
R(T,\varepsilon,L) =\varepsilon ^{\frac{(2+d)(\gamma-1)}{2\gamma}} 
R(T/\varepsilon^{2-\frac{(2+d)(\gamma-1)}{2\gamma}},1, L T^{\frac{\gamma-1}{2 \gamma -\gamma  d+d+2}}). 
\end{equation}
Alternatively, for large sizes $L T^{\frac{\gamma-1}{2 \gamma -\gamma  d+d+2}} \gg 1$, we can write
\begin{equation}
R(T,\varepsilon) = T^{\frac{(\gamma -1) (d+2)}{2 (\gamma +1)-(\gamma -1) d}} 
{\mathcal F}(\varepsilon^{2-\frac{(2+d)(\gamma-1)}{2\gamma}}/T)
\end{equation}
with ${\mathcal F}(x)$ a master function,
\begin{equation}
{\mathcal F}(x) \equiv x^{\frac{(\gamma -1) (d+2)}{2 (\gamma +1)-(\gamma -1) d}} R(1/x,1,{\infty}) 
\end{equation}
At $T=0$ is easy to see that the velocity at the depinning 
transition in this family of periodic potentials 
is $v \sim (f-f_c)^\beta$ with 
$\beta=1-1/\gamma$, 
since the problem reduces to the particle 
case~\cite{purrello2017}. 
If we use that $\psi \approx \beta/\alpha$ we find 
the thermal rounding exponent
$\psi =\frac{2 \gamma -2}{2 \gamma -\gamma  d+d+2}$.
In $d=1$ we get in particular $\psi =\frac{2 \gamma -2}{\gamma +3}$, 
so for $\gamma=2$ we have $\psi =\frac{2}{5}$. 

\begin{figure}
\includegraphics[width=8cm,clip=true]{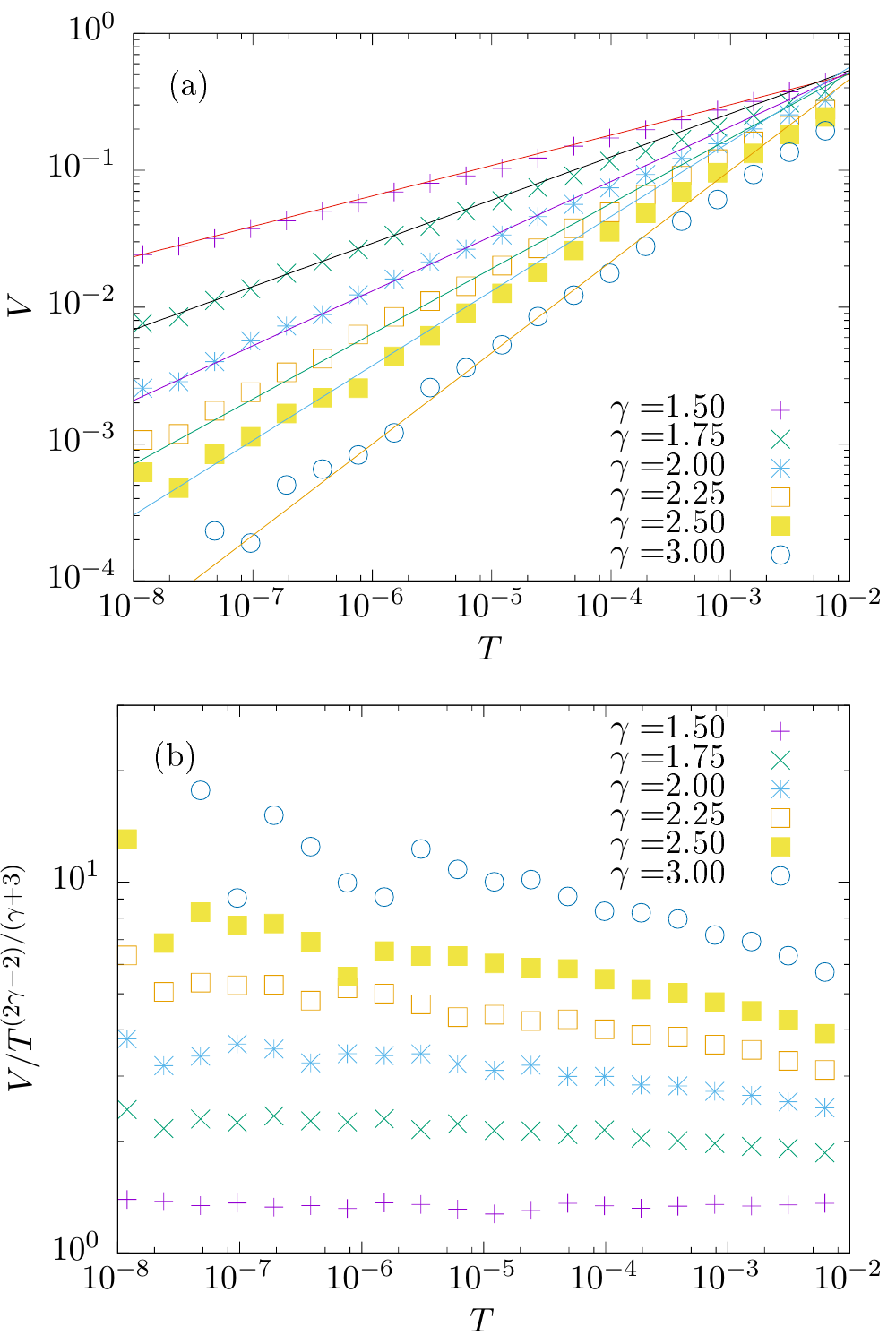}
\caption{(a) Velocity vs temperature at the depinning 
transition for an elastic string ($d=1$) of size 
$L=132768$ in a periodic potential characterized 
by the normal form exponent $\gamma$. 
Larger $\gamma$ imply shallower force minima 
(see text). Lines display the power-law behavior   
$v \sim T^{\psi(\gamma,d)}$, where
$\psi =\frac{2 \gamma -2}{2 \gamma -\gamma  d+d+2}$.
(b) Same data but highlighting the corrections to the 
power-law by plotting $v /T^{\psi(\gamma,d=1)}$ vs $T$. 
\label{fig:vvsTvsgamma}
}
\end{figure}
In Fig. \ref{fig:vvsTvsgamma} we compare with $v$ data 
at $f_c=1$, vs temperature $T$. As can be appreciated 
in Fig. \ref{fig:vvsTvsgamma}(a)
the ansatz $\psi=\frac{2 \gamma -2}{2 \gamma -\gamma  d+d+2}$ 
works reasonably, but corrections to the pure law 
scaling manifest already for temperatures $T>10^{-4}$ 
(in units of the microscopic energy scale which we 
have set to unity). Interestingly, 
as shown in Fig. \ref{fig:vvsTvsgamma}(b), these corrections
are accentuated for larger $\gamma$, corresponding 
to shallower bottlenecks $h^{\gamma}-\epsilon$ around 
$h=0$. As we have discussed for $\gamma=2$ case, 
logarithmic corrections are originated in the wide distribution of the
nucleation times. The enhancement of corrections for 
increasing values of $\gamma$ indicates that this distribution 
becomes wider as $\gamma$ increases.

\section{An independent justification for the existence of logarithmic corrections at $f=f_c$}

In the main part of this work, we have analyzed the velocity of an interface in a washboard potential
at finite temperatures. When the driving force $f$ is larger than the critical force $f_c$, the limiting velocity as $T\to 0$ corresponds to the athermal limit $v\sim (f-f_c)^\beta$, with $\beta=1/2$.
In the case in which $f<f_c$, and for $d=1$ the velocity vanishes in the limit $T\to 0$ following an activation scaling
that includes an Arrhenius factor of the form $\sim \exp(-C|f-f_c|^{5/4}/T)$. We then analyzed the behavior of the velocity right at the critical point, namely $f=f_c$, showing that this velocity has a dominant power law term in $T$, plus some logarithmic correction.

In the present Appendix, we want to show formally why this kind of logarithmic correction appears naturally  in this problem.  To this end, we pose the following mathematical problem which however is clearly related to the physical problem we have studied. Suppose we consider a function $v$ (to be associated to the velocity of the interface) as a function of $x\equiv f_c-f$, 
and temperature $T$. Suppose we know that $v(x,T)\to |x|^\beta$ as $x\to-\infty$\footnote{In the real problem of the interface, the form $v\sim |x|^\beta$ does not really hold in the $x\to-\infty$ limit. However, since we are interested in a small neighborhood of $T=0$, there will always be a range for which $|x|$ is ``sufficiently large'', and $v\sim |x|^\beta$ holds}, for any $T$. Also, suppose that as $x\to+\infty$ we know $v(x,T)\simeq T^{A}\exp(-x^\alpha/T)$ (compare with Eq. (\ref{Bl}); for simplicity we do not consider here the possibility of a pre-exponential power of $x$). The problem we pose is to find a consistent family of functions $v(x,T)$ for all $x$, satisfying the previous limiting forms. Once the family of functions has been found, we are mainly interested in the thermal rounding function $v(x=0,T)$.

\begin{figure}
\includegraphics[width=8cm,clip=true]{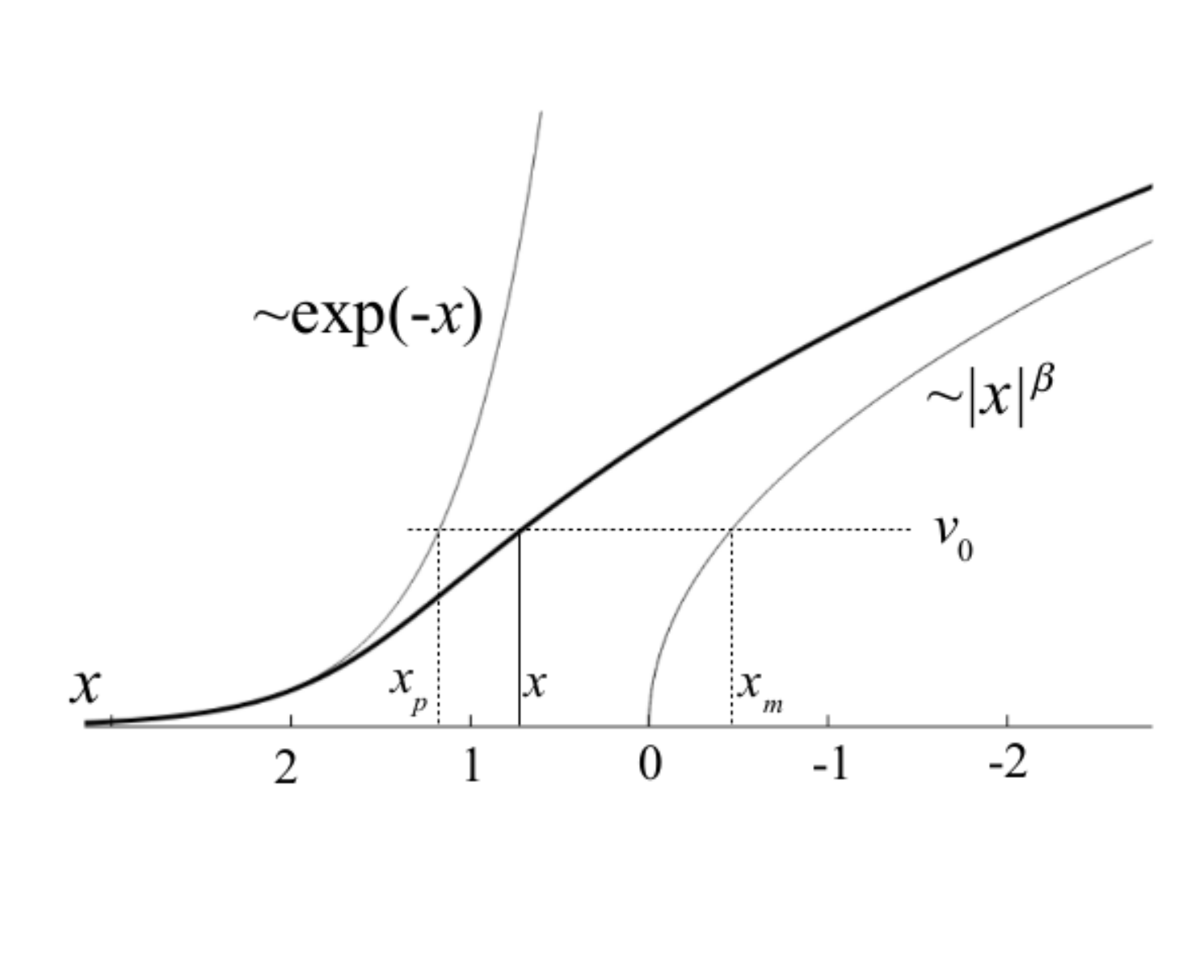}
\caption{Sketch of a function of which we now its limiting forms $v_p$ and $v_m$ as $x\to \pm \infty$. A possible interpolating function $v(x)$ for all $x$ consists in fixing some value $v_0$, and choosing $x$ such that $x\equiv v_p^{-1}(v_0)+v_m^{-1}(v_0)$
\label{fig:apend}
}
\end{figure}

In this very general form there will be of course many different solutions to the problem. Our goal here is to show how in  one possibly (arguable one of the simplest)  solutions that can be obtained, a logarithmic correction in the thermal rounding function appears, which is originated in the exponential form of the limiting function for $x\to+\infty$. In fact, in Fig. \ref{fig:apend} we can see the formal problem we are posing. There we see plotted with thin lines the two limiting functions for $x\to\pm\infty$. We call them $v_p$ and $v_m$ for concreteness. The thick line is an example of a possible function interpolating between these two limits. One simple form of analytically obtaining one such interpolation function consists in the following:  Fixing a generic value $v_0$ of $v$ ($v_0>0$) we obtain the two points $x_p$ and $x_m$ such  that 
$x_p=v_p^{-1}(v_0)$, $x_m=v_m^{-1}(v_0)$. Then we define the interpolating function $v(x)$, choosing $x=x_p+x_m$. Being more explicit, we define the inverse function $v^{-1}$ as 
\begin{equation}
v^{-1}(v_0)= {v_p^{-1}}(v_0)+{v_m^{-1}}(v_0) 
\end{equation}
Using the explicit forms $v_p(x)\equiv T^{A}\exp(-x^\alpha/T)$, $v_m(x)\equiv |x|^\beta$ we obtain
\begin{equation}
x\equiv v^{-1}(v_0)=v_0^{1/\beta}+[-T\log(v_0 T^{-A})]^{1/\alpha}
\label{fm1}
\end{equation}
Let us note the following. As the two terms of this definition have positive derivatives, the function $v^{-1}$
can be re-inverted to obtain a single valued 
$v_0(x)\equiv v(x)$
function.  In addition, when 
$v_0\to\infty$, the first term in (\ref{fm1}) dominates, whereas when $v_0\to 0$ is the second term that dominates. Then the two limits of the function $v(x)$ are satisfied.

Expression \ref{fm1} cannot be inverted analytically in general, but we can advance further considering the thermal rounding function, namely the case $x=v^{-1}(v_0)=0$. We obtain
\begin{equation}
0=v^{1/\beta}+[-T\log(vT^{-A})]^{1/\alpha}
\end{equation}
or 
\begin{equation}
v^{\alpha/\beta}=-T\log(vT^{-A})
\label{vdet2}
\end{equation}
This expression can be considered as it stands, as an implicit form of the thermal rounding curve. Alternatively we can get an explicit form by working iteratively: 
Taking into account the slow variation of the logarithmic factor, we can solve it first by considering the log term is a constant,  to obtain
$v\simeq (CT)^{\beta/\alpha}$,
and then using this zero-order approximation inside the log in Eq. (\ref{vdet2}), to obtain:
\begin{equation}
v\simeq  T^{\beta/\alpha}\left (-\log\left ((CT)^{\beta/\alpha}T^{-A}\right)\right)^{\beta/\alpha}
\end{equation}

Either from this approximate expression, or from the general solution (Eq. (\ref{vdet2})), we can see that in the particular case in which $A=\beta/\alpha$, the logarithmic correction vanishes and the solution is $v\sim T^{\beta/\alpha}$. This is just the case in which the activation form of the function we are looking for in the limit $x\to\infty$ is compatible (in the sense discussed at the end of section III B) with the form $|x|^\beta$ in the limit $x\to\-\infty$. Except in this particular case, logarithmic effects are expected.

We see that the appearance of a logarithmic correction seems to be a very general result associated to: {\em i)} the impossibility of fitting the two limiting expressions for $x\to\pm\infty$ with a single scaling relation (as the one in Eq. \ref{naivescaling}); {\em i)} the exponential activation form for $x\to-\infty$, whose inversion is  responsible for the logarithmic factor in the thermal rounding law.

\bibliography{biblio} 

\end{document}